# Free-electron laser-based extended wide-field mid-infrared photothermal imaging for biomedical and microplastic analysis

*Anooj Thayyil-Raveendran[1], Subham Adak[1], Artem Shydliukh[1], Natalja Redinger[2],*

*Matthias Hauptmann[2,3], Ulrich E. Schaible[2,3,4], Anna Mühlig[5], J. Michael Klopf[6],*

*Orlando Guntinas-Lichius[5], Jürgen Popp[7] and Christoph Krafft[1*]*

[1] Leibniz Institute of Photonic Technology, Member of Leibniz Research Alliance Leibniz Health Technologies, Member of the Leibniz Center for Photonics in Infection Research, Albert-Einstein-Str. 9, 07745 Jena, Germany

[2] Research Center Borstel, Leibniz Lung Center, Program Area Infections, Dept. Cellular Microbiology, Borstel, Germany;

[3] Leibniz Research Alliance INFECTIONS, Borstel, Germany;

[4] German Center for Infection Research (DZIF), Partner Site Hamburg-Lübeck-Borstel-Riems, Borstel, & Biochemical Microbiology and Immunochemistry, University of Lübeck, Lübeck, Germany;

[5] Jena University Hospital, Department of Otorhinolaryngology, 07747 Jena, Germany

[6] Institute of Radiation Physics, Helmholtz-Zentrum Dresden Rossendorf, Bautzner Landstr. 400, 01328 Dresden, Germany.

[7] Friedrich-Schiller-University Jena, Institute of Physical Chemistry, Member of Leibniz Research Alliance Leibniz Health Technologies, Member of the Leibniz Center for

*Corresponding author Christoph.Krafft@leibniz-ipht.de

Photonics in Infection Research, Helmholtzweg 4, 07743 Jena, Germany

**ABSTRACT**: Wide-field mid-infrared photothermal (MIP) imaging offers rapid label-free chemical contrast for biomedical and polymer analysis. However, its field of view (FOV) is limited by the pulse intensity of conventional infrared lasers. Here, we present a wide-field MIP microscope that uses a high-power free-electron laser (FEL) rather than a quantum cascade laser (QCL) as the pump source to achieve a substantially larger FOV. Both implementations use counter-propagating beam paths with a 450 nm LED as the probe source and a CMOS camera that records images using a virtual lock-in detection scheme. QCL nanojoule pulse energies enables FOV of around 45 micrometers for wide-field MIP imaging with a sub-micrometer resolution for polystyrene beads, Mycobacterium tuberculosis infected fixed tissues, and laryngeal cancer cryosections. IR spectra in the range of 1000–1800 wavenumbers can be reconstructed by tuning the QCL. FEL pulse energies of up to microjoules expand the FOV by a factor of nearly 20 as demonstrated by wide-field MIP imaging of polystyrene beads, single cells, and murine brain tissue. We discuss current challenges and further improvements to implement high-power IR lasers for wide-field MIP imaging with even larger FOVs in the context of biomedical research and diagnostics.

*Corresponding author Christoph.Krafft@leibniz-ipht.de

# Introduction

Infrared (IR) spectroscopy is a label-free method of probing molecular vibrations, enabling the identification of chemicals in fields ranging from biology to materials science[1]. The mid-IR range (2.5–25 µm or 4000 – 400 cm$^{-1}$) is of particular relevance, as it covers the fundamental vibrational modes that provide a molecular "fingerprint" of materials. While IR spectroscopy directly probes molecular vibrations through absorption, Raman spectroscopy provides complementary vibrational information via inelastic light scattering. Raman-based techniques routinely achieve sub-micron spatial resolution using visible excitation lasers and high numerical aperture objective lenses. Coherent Raman variants such as stimulated Raman scattering (SRS) and coherent anti-Stokes Raman scattering (CARS)[2-4] coupled to laser-scanning microscopes offer video-rate imaging with microsecond dwell times per pixel. However, Raman techniques are limited by their inherently low scattering cross sections, which restricts their sensitivity[5]. In contrast, mid-IR absorption-based methods boast absorption cross sections that are approximately eight orders of magnitude larger and therefore provide a significantly more sensitive alternative for label-free chemical imaging. Direct mid-IR techniques, including wide-field FTIR microscopes with focal plane arrays (FPA)[6,7] and quantum cascade laser (QCL)-based microscopes with microbolometer arrays[8-11], enable high-speed imaging. FTIR microscopes with 128×128 FPAs typically achieve fields of view (FOV) of ~700×700 µm² at a pixel size of 5.5 µm, while tunable QCLs coupled to array detectors improve both speed and FOV. This allows for FOV of 650×650 μm² at a pixel size of 1.3×1.3 μm2 or 2×2 mm² at a pixel size of 4.3×4.3 μm2. However, the lateral resolution of direct wide-field IR imaging techniques depends on the wavelength and numerical aperture according to Abbe's diffraction equation, ranging from approximately 5 to 2.8 μm in the fingerprint region (1000–1800 cm$^{-1}$, corresponding to 10–5.5 μm wavelengths). Indirect IR detection schemes which are pumped by the absorption of pulsed IR laser radiation, induce photothermal effects in specimens, resulting in local and temporary changes to the refractive index and volume[12]. These photothermal effects can be probed using a visible laser enabling detection by light microscopes with submicron resolution depending on the probe laser wavelength[13]. The initial implementation of mid-IR photothermal imaging (MIP), also known as optical photothermal infrared (OPTIR), used point scanning for discrete-wavenumber chemical mapping with a motorized stage and a dwell time of milliseconds per pixel for example, for cell biological and pharmaceutical studies[14,15]. Although point scanning OPTIR enables chemical mapping at high spatial resolution, it was initially limited to approximately 1000 points per second taking more than 15 minutes to capture a 1-megapixel image. Recently, a laser-scanning OPTIR approach improved imaging speed by a factor of approximately 30 [16].

*Corresponding author Christoph.Krafft@leibniz-ipht.de

Wide-field implementations of MIP microscopic imaging using both expanded IR pump lasers and visible probe sources, and large area and high-speed CMOS camera sensors were summarized[17]. Wide-field MIP sensing enabled imaging of polymers and single cells within a 36×36 μm² FOV at speeds up to 1250 frames per second (fps), providing sub-micrometer spatial resolution using an OPO-based mid-IR source (output power up to ~30 mW) in the fingerprint region (1176–1786 cm$^{-1}$)[18]. The IR absorption-induced photothermal expansion with a QCL excitation can also be detected with a wide-field interferometric approach that exceeded FTIR imaging ten-fold in coverage, four-fold spatial resolution and spectral consistency by mitigating anomalous scattering effects in the fingerprint region[19]. An improved wide-field MIP detection was developed by coupling with a quantitative phase microscope[20]. This detection strategy provided a two-order-of-magnitude improvement in signal-to-noise ratio compared to earlier wide-field MIP methods, enabling live-cell imaging beyond video rate using an OPO source at in the high-wavenumber region (2600 - 3450 cm$^{-1}$) with pulse energies up to 10 μJ. Wide-field mid-infrared heterodyne imaging has also reached acquisition speeds of up to 4000 fps using an EC-QCL (peak power up to 160 mW) operating in the silent region (2015–2220 cm$^{-1}$)[21]. Wide-field fluorescence-detected mid-infrared photothermal imaging (F-MIP) probes the temperature-induced quenching of fluorophores (around 1% per Kelvin), resulting in around 100 times greater sensitivity than scattering-based MIP methods when driven by a QCL (output power from 5 to 15 mW) from 1000 to 1886 cm$^{-1}$ [22]. F-MIP was also implemented on a commercial OPTIR platform, allowing for wide-field imaging of intrinsic fluorophores in biological samples like diatoms, plant tissues, and microalgae within a ~66×66 μm² FOV[23]. Despite the development of several wide-field MIP imaging systems, their fields of view remain significantly restricted by the output power of mid-IR excitation sources like OPOs and QCLs. The usable imaging area is usually limited to diameters of less than 100 micrometers because the mid-IR average intensity[24], typically below 20 mW [25], is insufficient to generate photothermal response across larger areas.

*Corresponding author Christoph.Krafft@leibniz-ipht.de

*Corresponding author Christoph.Krafft@leibniz-ipht.de

In this work, a wide-field MIP microscope was developed that consisted of the pulsed and tunable QCL as IR pump source, a pulsed, visible LED as probe source, a CMOS camera and a function generator. A virtual lock-in detection scheme synchronized registration of hot and cold frames with QCL emission on and off, respectively. The setup was applied to wide-field MIP imaging of polystyrene (PS) beads, formaldehyde-fixed *Mycobacterium tuberculosis* infected murine lung tissue sections, and human laryngeal cancer sections on CaF$_2$ windows. IR spectra were reconstructed from hyperspectral data sets by tuning the IR laser wavenumber and sequential registration of MIP images. As the FOV was limited to around 45 μm diameter by the intensity of the QCL-IR pulses, the wide-field MIP microscope was installed at a free-electron laser (FEL) facility for more intense IR excitation and effective expansion of FOV. Utilizing its broad tunability and high pulse intensities, the FEL is a versatile radiation source across several disciplines[26-30]. For the first time, this work describes extended wide-field FEL-MIP imaging from PS beads, single cells and tissue sections and compares the resulting images with wide-field QCL-MIP images. These experiments demonstrate that wide-field

*Corresponding author Christoph.Krafft@leibniz-ipht.de

MIP imaging enables fast, label-free screening of large FOVs offering significant advantages for cancer diagnostics, infectious disease research and microplastic analysis.

## Results

### Widefield QCL-MIP microscope

The optical configuration of the widefield QCL-MIP microscope was realized in counter propagation mode with the QCL pump beam at the bottom and the LED probe beam at the top (Fig. 1a). The detection path is based on a widefield reflection microscope. Details of the components are compiled in the experimental section. To achieve uniform illumination at the sample plane, a plano-convex lens L1 focuses pulsed LED beam at the rear focal plane of the objective lens via a beam splitter. Using a 40×/0.6 NA microscope objective lens and a 450 nm wavelength light source, the lateral resolution is expected to be near 375 nm according to the Abbe limit. A tube lens L2 focuses the reflected light from the sample onto a CMOS sensor after it passes through the same objective lens and beam splitter. The IR beam was focused on the sample by a ZnSe plano convex lens L3. A rectangular waveform generated by the function generator synchronizes QCL pulses, LED pulses and camera readout. The alternate IR-on (hot frame) and IR-off (cold frame) images were then registered by the CMOS camera (depicted in Fig. 1b). Photothermal contrast originates from the difference between hot and cold frames, resulting from localized heating that produces noticeable alterations in scattering and refractive index, which become apparent when the two frames are compared.

*Corresponding author Christoph.Krafft@leibniz-ipht.de

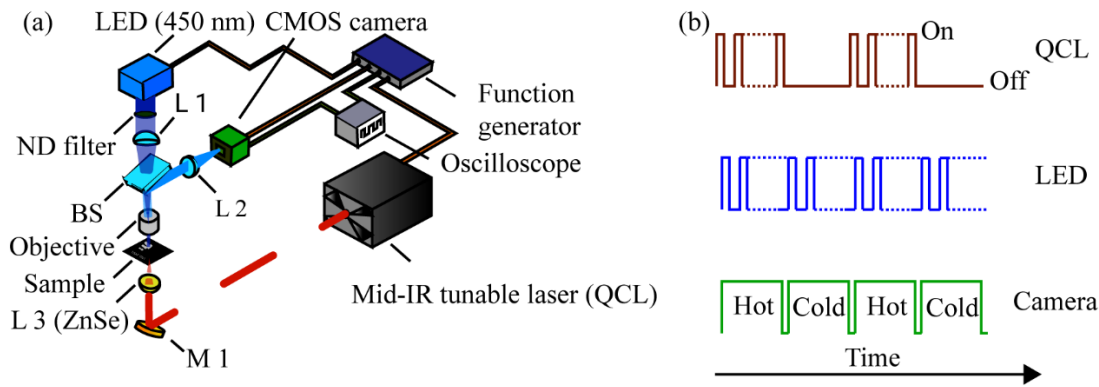

**Figure 1. Optical and electronic set-up of widefield QCL-MIP microscope.** (a) Configuration to capture MIP images by the pump-probe imaging principle with the function generator triggering QCL, LED and CMOS camera. (b) Representation of pulses from QCL, LED, and camera (dotted line indicating several pulses in between) to capture alternate hot and cold frames.

**Wide-field QCL-MIP imaging of polystyrene beads**

We compare wide-field QCL-MIP images of 10 μm diameter polystyrene beads with commercial OPTIR instrument. The wide-field QCL-MIP image at 1450 cm$^{-1}$ (Fig. 2b) showed a high chemical contrast in a FOV of approximately 45 μm diameter (ca. 1.6×10$^3$ μm2) whereas the image at 1660 cm$^{-1}$ showed low off-resonance contrast (Fig. 2c). Analogous to the visible image, the wide-field-MIP image shows maximum intensities at the margin and in the center, and minimum intensity in a ring between the maxima. The "bright center–dark ring–bright edge" pattern may originate from lensing effects of the bead's spherical shape. Scattering and local field effects at the edges and center increase the photothermal signal in these areas, with the intermediate region displaying weaker contrast as a result of reduced scattering.

*Corresponding author Christoph.Krafft@leibniz-ipht.de

For comparison, QCL-MIP images were collected in scanning mode using the commercial Mirage-R instrument in reflection mode. Details are given in the experimental section. The visible microscope image with the 40× Cassegrain objective lens shows triangle features in the center of PS beads which is due to reflective mirror of such lenses (Fig. 2d). Scanning a 45 µm×45 µm area with a step size of 150 nm at scan speed of 45 µm/s took approximately 440 seconds which is more than two orders of magnitude slower compared to 1-2 seconds total time of the wide-field version. The step size was comparable to the pixel size of the CMOS sensors after 2×2 binning. Similar to the wide-field-MIP, visible and IR contrast at 1450 cm$^{-1}$ are maximum in the center of the PS beads (Fig. 2e), followed by a dark ring and a weak contrast at the margin. The off-absorption contrast at 1660 cm$^{-1}$ shows weak features in the center (Fig. 2f).

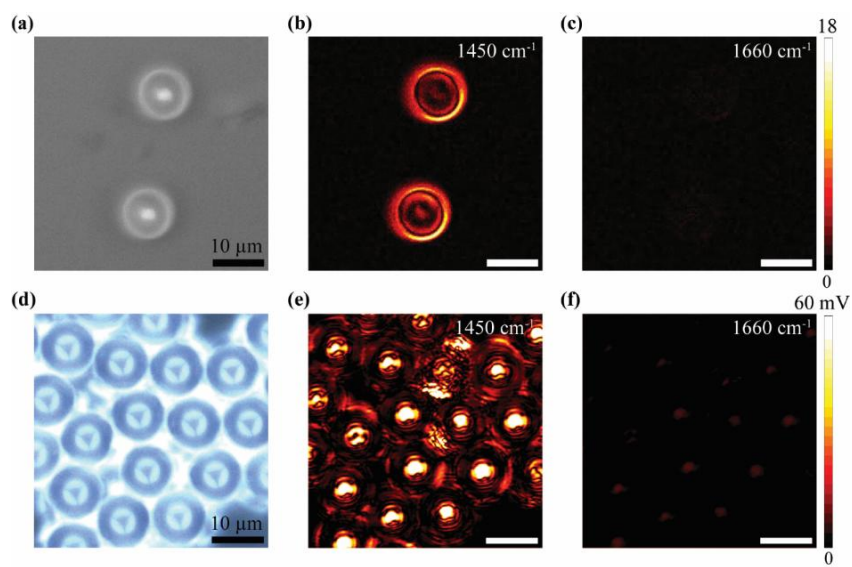

**Figure 2. Wide-field QCL-MIP images of 10 µm polystyrene beads.** (a) Visible image from wide-field setup at 40× magnification. (b) MIP image at 1450 cm-1 with 3.8 mW average IR power and (c) at 1660 cm$^{-1}$ with 7.5 mW average IR power. (d) Visible image

*Corresponding author Christoph.Krafft@leibniz-ipht.de

from Mirage-R instrument with 40× Cassegrain objective. (e) OPTIR image at 1450 cm$^{-1}$ (IR power – 0.6 mW) and (f) at 1660 cm-1 using the Mirage-R microscope at 150 nm step size, 45 µm/s scan speed, 0.4 mW IR power and 5.4 mW probe power at 785 nm. Same intensity scale was selected in (b), (c) and (e), (f).

**Wide-field QCL-MIP imaging of *M. tuberculosis* infected murine lung tissue**

Wide-field MIP images were collected from different areas of lung tissue sections from *M. tuberculosis*-infected susceptible C3HeB/FeJ mice treated with bedaquiline on a CaF$_2$ window (Fig. 3a-c). C3HeB/FeJ mice are considered susceptible to *M. tuberculosis* infection developing necrotic granulomas, an inflammatory tissue reaction mimicking human tuberculosis lesions. 500 hot and cold frames were obtained from lung tissue at 400 fps on average at 1655 and 1740 cm$^{-1}$ that correspond to the protein amide I band and lipid-associated C=O stretching vibrations, respectively, and provide insight into protein and lipid distribution. Within the infected tissue section, we identified cells with noticeably increased lipid contrast (1740 cm$^{-1}$) compared to the surrounding tissue, indicating lipid-rich materials such as lipid droplets. The intensities of the protein contrast are similar in the lipid-rich cell and the surrounding tissue. We suggest that the obtained image of the lipid-rich cell show a so-called foamy macrophage, typically associated with active *tuberculosis* granulomas.

*Corresponding author Christoph.Krafft@leibniz-ipht.de

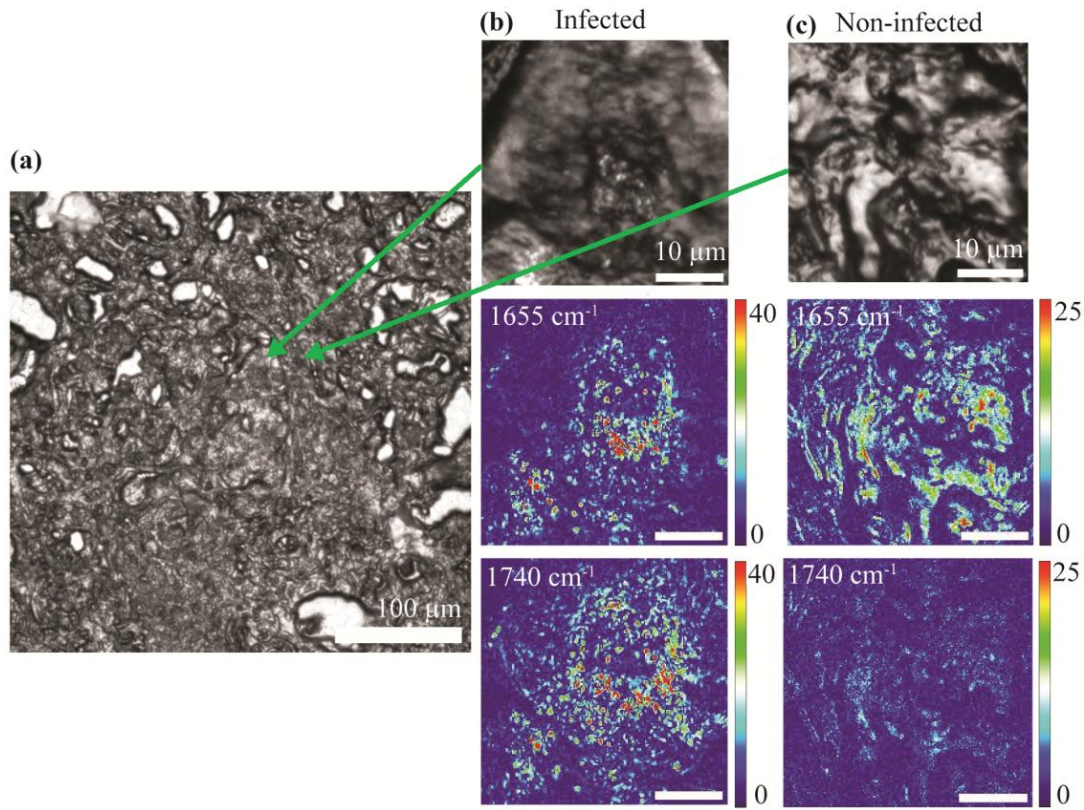

**Figure 3. QCL wide-field MIP imaging of a *M. tuberculosis* infected murine lung tissue.** (a) Microscopic overview image with 450 nm illumination. (b) Zoomed microscopic image of a lipid-rich cell, wide-field MIP images at 1655 and 1740 cm$^{-1}$. (c) Zoomed microscopic image of surrounding, wide-field MIP images at 1655 and 1740 cm$^{-1}$.

**Wide-field QCL-MIP imaging of human cancer tissue section**

Wide-field MIP images were collected from a human larynx tissue section containing both cancerous and non-cancerous regions mounted on a CaF$_2$ window. An optical microscopic image and wide-field MIP images of cancer tissue at 1660 cm$^{-1}$ representing the amide I vibration of proteins and at 1740 cm$^{-1}$ representing the C=O vibration of lipids are shown in Fig. 4a, b, and c, respectively. Microscopic and wide-field MIP images at these same wavenumbers were also collected from a non-cancerous region

*Corresponding author Christoph.Krafft@leibniz-ipht.de

(Fig. 4d, e and f) for comparison. The diameter of the area from which IR contrast was obtained was estimated as 45 µm. The intensities of the protein images are similar, but their distributions are different. The red spots might correlate with the regions of higher protein content in the cancerous area. Due to altered cell structure in tumor cells certain proteins like kinases and other enzymes can appear in higher concentrations compared to non-cancerous cells. The cracks in the microscopic image (Fig. 4a) are assigned to cryo artefacts. The cracks are also resolved in the wide-field MIP image (Fig. 4b and 4c) which might be due to scattering effects. The overall lipid contrast is less intense in the cancer than in the non-cancer MIP image (Fig. 4c and f).

To validate the different lipid to protein ratios, IR spectra of cancerous and non-cancerous areas were reconstructed from hyperspectral wide-field MIP images and compared with IR spectra acquired using commercial OPTIR and FTIR instruments. Hyperspectral wide-field MIP images were collected by tuning the QCL from 980 to 1760 $cm^{-1}$ at an interval of 5 $cm^{-1}$. Collecting series of 157 images took ca. 12 min total acquisition time. To reconstruct the IR spectra, the average pixel intensity inside the ROI with high IR intensity (blue and red boxes) and outside the ROI (green boxes) with low IR intensity was determined from the wide-field MIP image, and the difference of both average intensities was calculated for each wavenumber. As the QCL emission varies in the spectral range and water vapor absorption from ca. 1400 to 1700 $cm^{-1}$ overlaps with the absorption of the biomolecules, division of the raw spectra by the power spectrum of the QCL (Fig. S1) gave MIP spectra (Fig. 4g) that coincided well with OPTIR and FTIR spectra of commercial instruments (Fig. 4h and i), respectively. Spectral markers were found in all spectra such as more intense bands near 1740 cm$^{-1}$ indicating elevated lipid content in non-cancer, and more intense bands near 1080 cm$^{-1}$ indicating elevated nucleic acid content in cancer tissue relative to the amide I bands near 1660 cm$^{-1}$ that was used for normalization. Data collection from slightly different positions explains small variations between wide-field MIP, OPTIR and FTIR spectra e.g. the ratio of the bands near 1740 relative to 1660 cm$^{-1}$.

*Corresponding author Christoph.Krafft@leibniz-ipht.de

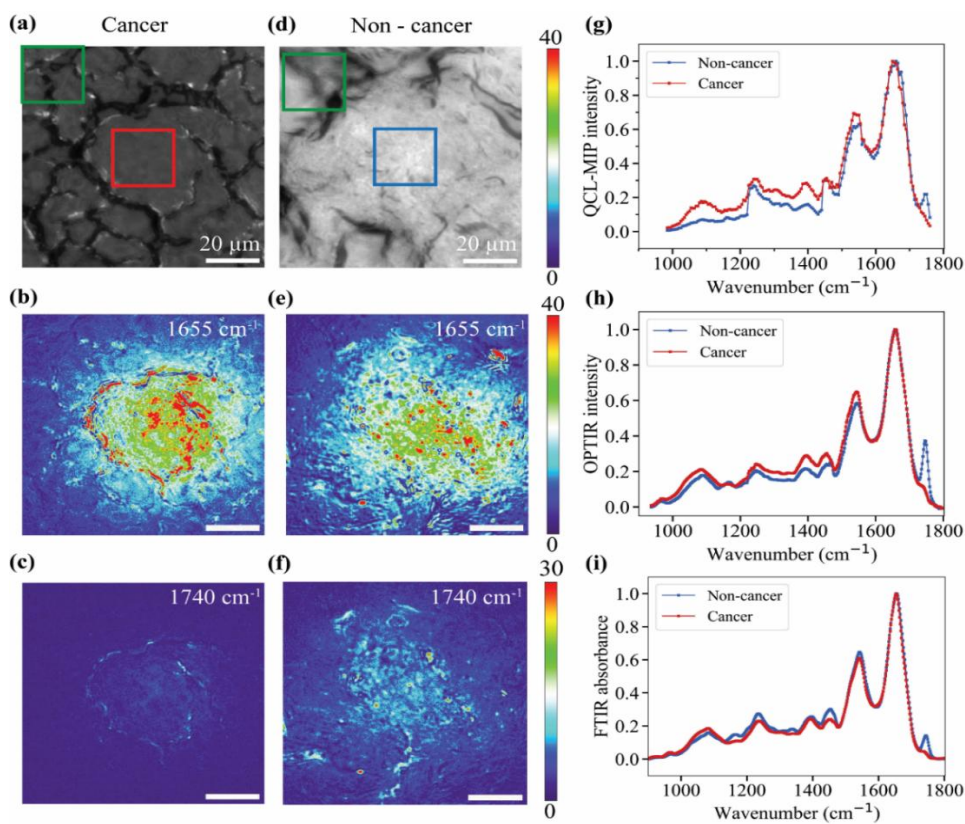

**Figure 4. Wide-field QCL-MIP imaging of a cancer tissue section.** (a) Microscopic images of cancerous laryngeal tissue area. (b) and (c) wide-field MIP images of the cancerous area at 1655 and 1740 cm$^{-1}$ (d) Microscopic images of non-cancerous laryngeal

*Corresponding author Christoph.Krafft@leibniz-ipht.de

tissue area. (e) and (f) wide-field MIP images of the non-cancerous area at 1655 and 1740 cm$^{-1}$ (g) Reconstructed wide-field MIP spectra, (h) single point commercial OPTIR spectra, and (i) FTIR spectra of cancer and non-cancer tissue. Red and blue boxes indicate regions from which the IR spectra were reconstructed from hyperspectral wide-field MIP images. Background was obtained from the region of the green boxes.

**Wide-field FEL-MIP microscope**

The FOV of the wide-field QCL-MIP microscope was limited by the energies of QCL pulses which were focused to a ca. 45 µm diameter spot by a short focal lens L3. FELs emit IR radiation at several orders of higher pulse and average intensities. Relevant parameters of the external-cavity (EC-) QCL (four-chip Mircat2400, Daylight) and the FEL FELBE are summarized in Table 1. FELBE consists of two oscillator FELs operating in the THz-MIR range (40–2000 cm$^{-1}$ / 250–5 µm)[31] and emitting picosecond pulses that are nearly transform-limited with a typical bandwidth of 1–2% at a fixed repetition rate of 13 MHz. With pulse energies reaching as high as a few µJ at some wavenumbers below 1000 cm$^{-1}$, FELBE delivers extremely high peak and average spectral brightness for a wide range of experiments. The pulse energy in the fingerprint range was up to 150 nJ. Furthermore, each FEL is continuously tunable over approximately a decade of spectral frequency/wavelength by adjusting the electron beam energy or the K-parameter of the undulator. The K-parameter of the FELBE FELs can be set even by the user by varying the gap of the permanent magnet undulator. This allows the user to easily take hyperspectral measurements over a wide spectral range (ca. ±25% of the wavelength).

*Corresponding author Christoph.Krafft@leibniz-ipht.de

|        | Repetition rate | Pulse length | Spectral bandwidth | Pulse Energy | Average power | Tuning range | Tuning speed/ |
|--------|-----------------|--------------|--------------------|--------------|---------------|--------------|---------------|
| EC-QCL | 0.1-2 MHz       | 20-1000 ns   | 0.5 cm$^{-1}$      | Few nJ       | 1-15 mW       | 934-1800, 2700-3000 cm$^{-1}$ | Up to 30,000 cm$^{-1}$/s |
| FELBE  | 13 MHz          | 1-25 ps*     | 1-2%, (~10 cm$^{-1}$ @ 1000 cm$^{-1}$) | Up to ~150 nJ | 300-2000 mW | 40-2000 cm$^{-1}$ | ** |

**Table 1. Specifications of EC-QCL and free-electron laser FELBE** (ELBE Center for High-Power Radiation Sources, Helmholtz-Centrum Dresden-Rossendorf, Germany). *Wavelength dependent. ** Rapid either continuously or step-wise (5-10 s/step) over a limited range (typically +/- 25% of λ).

Figure 5 shows the coupling of our wide-field MIP microscope to FELBE. Step attenuators and the diameter of the iris aperture adjusted the power of the IR beam that was focused via an off-axis OAP1 parabolic mirror (focal length 20 cm) and mirror 1 onto the sample. The horizontal mounting of OAP1 required additional mirrors M2 to M5 for beam steering. An optical chopper between OAP2 and OAP3 controlled the on and off periods of the FELBE beam to register hot and cold image frames. The beam at 13 MHz was chopped at 50 Hz to record 100 hot and cold frames. The chopping frequency, image acquisition speed of a CMOS camera, and LED pulse modulation were controlled by a function generator.

*Corresponding author Christoph.Krafft@leibniz-ipht.de

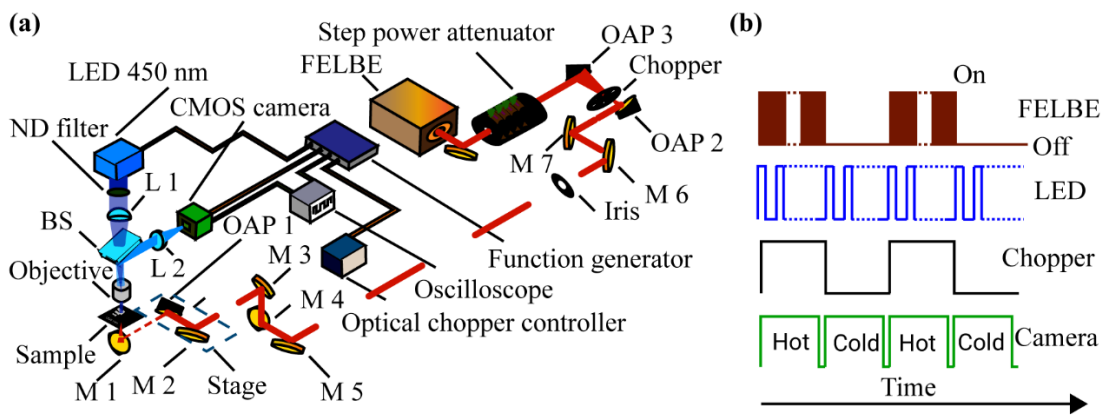

**Figure 5: Optical and electronic set-up of wide-field FEL-MIP.** (a) Configuration of optical and electronic components to capture MIP images by pump-probe imaging principle with the optical chopper and function generator triggering FELBE, CMOS camera and LED. (b) Representation of pulses from FELBE, LED, camera and optical chopper (dotted line indicating several pulses in between) to capture alternate hot and cold frames.

**Wide-field FEL-MIP imaging of polystyrene beads**

Wide-field FEL-MIP images were collected from a single FOV of 10 μm PS beads on a 1 mm thick $CaF_2$ window (Fig. 6a). This experiment demonstrated for the first time wide-field FEL-MIP imaging over an extended FOV, taking advantage of the FEL's high intensity to significantly increase the illumination area compared with wide-field QCL-MIP systems. By altering the undulator gap (K-parameter), the FEL was tuned to 1660 $cm^{-1}$ (off resonance) and 1450 $cm^{-1}$ ($CH_2$ deformation vibrations). While the off resonant image exhibits minimal contrast (Fig. 6c), the image at 1450 $cm^{-1}$ (Fig. 6b) shows an excellent photothermal contrast. The IR beam of the FEL has an approximately Gaussian profile with an elliptical shape which is depicted in Fig. 6b. The extension of the major axis was ca. 225 μm which is ca. five times the width of the wide-field QCL-MIP image. Fig. 6d shows a magnified view of the box in Fig. 6b. The minima in the center are due to saturation of the LED probe pulses at the CMOS camera. Hyperspectral images were measured by tuning the laser from 1400 to 1510 $cm^{-1}$ using wide-field FEL- and QCL-MIP at 15 and 10 $cm^{-1}$ intervals (Fig. 6e). Wide-field QCL- and FEL-MIP images at each wavenumber are shown in supplementary Fig. S2. The spectra were reconstructed by (i) calculating the average intensity of 5 μm areas at three different positions in the wide-field MIP images at each wavenumber followed by (ii) scaling the intensity based on the IR power output at each wavenumber. The normalization of QCL spectra was already described in the context of Fig. 4. The bandwidth of the FEL beam was estimated to be near 16 $cm^{-1}$. For comparison an average IR spectrum of three beads is included in Fig 6e at a 2 $cm^{-1}$ interval collected with the commercial OPTIR instrument which performed an automated background normalization. The IR spectra obtained with the OPTIR system and wide-field QCL excitation match well, while the spectrum measured with FEL excitation shows broader bands, consistent with its inherently broader bandwidth (see Table 1). Spectral reconstruction for the beads was carried out using the same approach as described for the cancer tissue section in Fig. 4. The modest band shift in the IR spectra around 1450 and 1490 $cm^{-1}$ is due to the difference in wavenumber intervals between hyperspectral images and the linewidth of QCL versus FEL pulses.

*Corresponding author Christoph.Krafft@leibniz-ipht.de

*Corresponding author Christoph.Krafft@leibniz-ipht.de

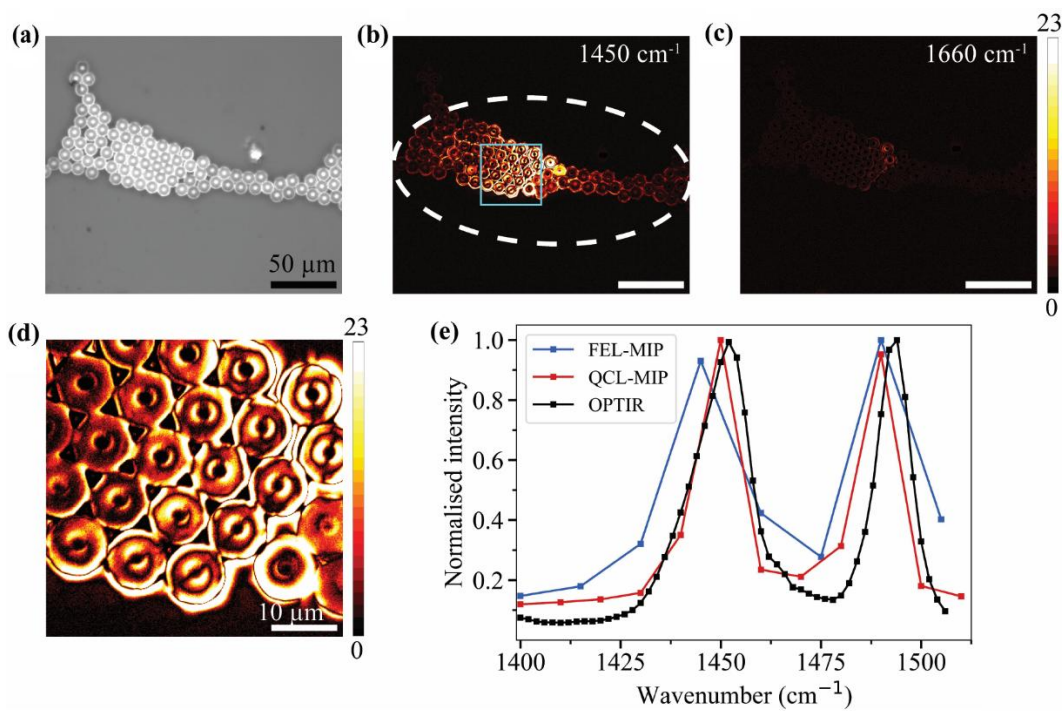

**Figure 6: Wide-field FEL-MIP imaging of PS beads.** (a) Visible image at 40× magnification. (b) Photothermal contrast at 1450 cm$^{-1}$ and (c) 1660 cm$^{-1}$ with 49 mW IR power and 45 µW LED intensity. (d) Enlarged view of green box in (b). (e) IR spectra of PS beads using FEL-MIP (blue), QCL-MIP (red) and OPTIR microscopes (black).

**Wide-field FEL-MIP imaging of a murine brain tissue section**

An extended wide-field FEL-MIP image of a mouse brain tissue section was compared with the wide-field QCL-MIP image. Wide-field FEL- and wide-field QCL-MIP images were collected at 100 fps, averaged over one second each, for intense protein band due to amide I vibrations at 1660 cm$^{-1}$, and for a lipid band due to CH$_2$ deformation vibrations at 1462 cm$^{-1}$. The visible images (Fig. 7a and c) are shown together with the weighted overlays (weight of 0.6 for the image at 1462 cm$^{-1}$ and 0.4 for the image at 1660 cm$^{-1}$) of wide-field FEL-MIP (Fig. 7b) and wide-field QCL-MIP (Fig. 7d).

*Corresponding author Christoph.Krafft@leibniz-ipht.de

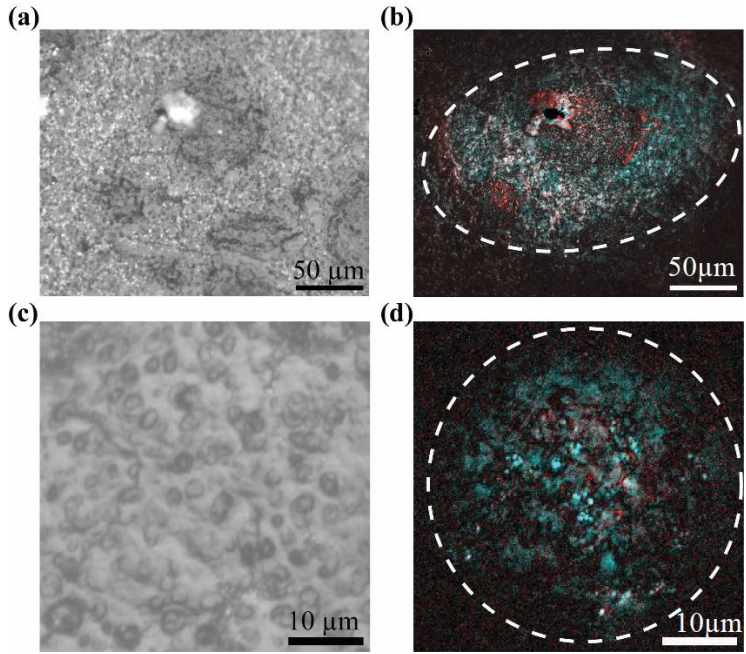

**Figure 7: Overlay images of protein and lipid distributions in mouse brain tissue sections using wide-field FEL- and QCL-MIP imaging.** (a) Visible image and (b) two color image representing proteins (cyan) and lipids (red) at 1660 and 1462 cm$^{-1}$, respectively, using the FEL-MIP microscope. (c) Visible image and (d) two color image representing proteins and lipids using QCL-MIP microscope.

The two-color images (proteins in cyan, lipids in red) enable to visualize their relative content. The wide-field FEL-MIP image encompasses an elliptical region of $3\times10^4$ µm$^2$ (with a major axis of 240 µm and a minor axis of 165 µm; Fig. 7b), roughly 20 times larger than the $1.6\times10^3$ µm$^2$ area visible in the wide-field QCL-MIP image (Fig. 7d), which effectively showcases the substantial field-of-view improvement achieved through the use of the FEL. The lipid features in the wide-field FEL-MIP image correlate with microcrystals in the visible microscope image. Such crystals are often observed in dried up brain tissue sections due to the high lipid content and the induction of oversaturating conditions[32,33]. Lipids constitute approximately 50-60% of the brain's dry weight while proteins make up 25-30%, though the exact percentages may vary between specific brain regions[34].

*Corresponding author Christoph.Krafft@leibniz-ipht.de

**Widefield FEL- and QCL-MIP for cell imaging**

The THP-1 cell line originally derived from a patient with acute monocytic leukemia served as model for white blood cells in an OPTIR study[35]. A small volume of THP-1 cell suspension was pipetted onto UV-grade 200 μm thick $CaF_2$ windows and dried. Multispectral wide-field FEL-MIP, QCL-MIP and scanning OPTIR images of THP-1 cells are shown at 1660, 1545, 1462 and 1350 cm$^{-1}$ representing amide I and II bands, $CH_{2/3}$ deformation vibrations, and off-resonance, respectively (Fig. 8).

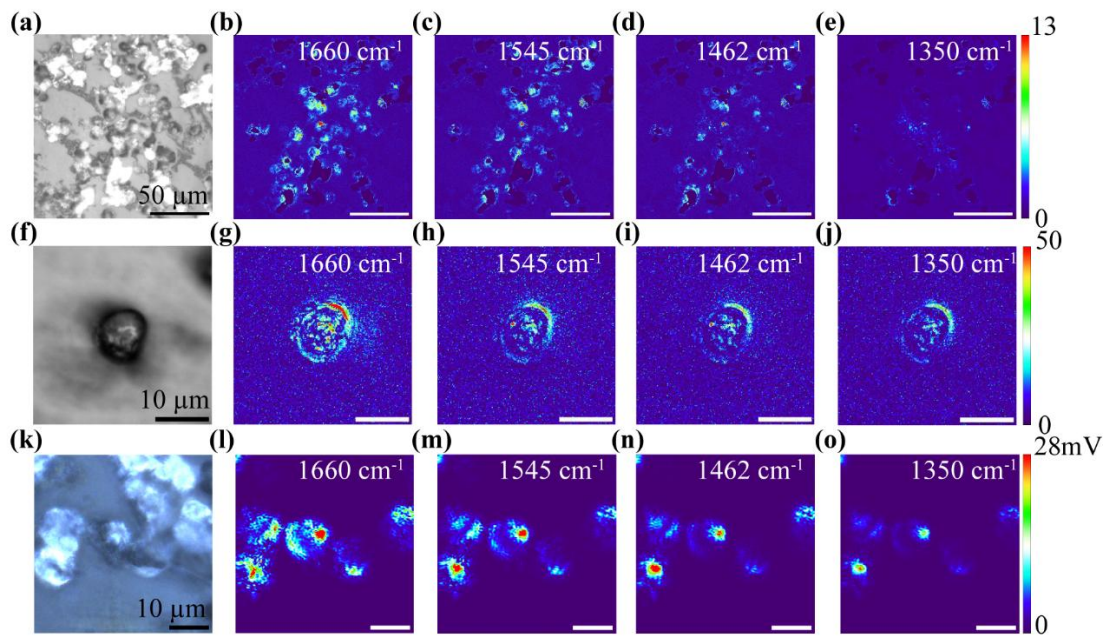

**Figure 8: MIP images of THP-1 cells at 1660, 1545, 1462 and 1350 cm-1 with 40× objective lenses.** (a) Visible image and (b-e) wide-field FEL-MIP images at IR power 168 mW, 205 mW, 216 mW and 190 mW, respectively; (f) Visible image and (g-j) wide-field QCL-MIP images at IR power 8.7 mW, 6.8 mW, 7.3 mW and 8.8 mW, respectively; (k) Visible image and (l-o) scanning OPTIR images.

*Corresponding author Christoph.Krafft@leibniz-ipht.de

The white regions in the visible images show buffer crystals that did not transmit IR radiation and appear dark in all FEL-MIP images (Fig. 8b-e). Cells showed the best contrast for the amide I band in all IR microscopic images (Fig. 8b, g, and l). The SNRs at 1660 cm$^{-1}$ in wide-field FEL-MIP, wide-field QCL-MIP, and scanning OPTIR images are calculated to be 14, 12, and 355, respectively. Two circular features are well resolved in all OPTIR images (Fig. 8l-o), even at the non-resonant wavenumber 1350 cm$^{-1}$ where low IR absorbance is expected according to Fig. 4g. The assignment of the features is not known yet.

**Discussion**

An in-house developed wide-field MIP microscope was first coupled to a QCL as IR pump source. The average intensity of less than 10 mW limited the FOV to a diameter of ca. 45 μm corresponding to ca. $1.6 \times 10^3$ μm$^2$. Second, the wide-field MIP microscope was coupled to FEL as IR pump source offering ca. 200 mW average intensity which gave an almost 20 time extended elliptical FOV of ca. $3 \times 10^4$ μm$^2$. The suitability was demonstrated for discrete wavenumber, multi- and hyperspectral imaging of PS beads, single cells and tissue sections. The photothermal response was registered using a CMOS camera at different frame rates and a time-gated pump-probe technique using a μs-pulsed blue LED as probe source. Wide-field MIP images resolved a broad range of specimens such as the distinct 3D shape of single 10 μm PS beads consisting of homogenous material, the cellular features of single THP-1 cells, and the complex protein and lipid distributions in *M. tuberculosis* infected murine lung, as well as cancer and non-cancer larynx, and murine brain tissue sections. Series of images upon tuning the wavenumbers enabled to reconstruct the IR spectra from wide-field MIP images. Both wide-field MIP microscope setups outperform the commercial scanning OPTIR instrument in terms of imaging speed. Measurements from FEL-MIP gave images with comparable SNR to the QCL-MIP which were, however, inferior to the SNR in the scanning OPTIR instrument. Table 2 summarizes the key performance characteristics of commercial QCL, wide-field QCL, and wide-field FEL-MIP systems. The parameters compared FOV, measurement time (including 100 frame averaging), and the respective advantages and limitations of each approach.

*Corresponding author Christoph.Krafft@leibniz-ipht.de

| Method | FOV | Typical measurement time | Properties |
|---|---|---|---|
| Scanning OPTIR | 45×45 µm | 440 s (150 nm step size at 45 µm/s) | Pros: better SNR, flexible FOV<br>Cons: scanning speed, hyperspectral imaging |
| Wide-field QCL-MIP | ⌀45 µm | <2 s (140 µm pixel size) | Pros with respect to scanning OPTIR: high speed imaging for discrete wavenumber, mosaic mode for extended FOV possible<br>Cons: low SNR and speed for reconstruction of spectra, FOV limited by low pulse intensity |
| Wide-field FEL-MIP | 240 µm×165 µm | <2 s (140 µm pixel size) | Pros with respect to QCL-MIP: extended FOV, tuning range towards low wavenumbers<br>Cons: FEL access |

**Table 2. Comparison of commercial QCL, wide-field QCL, and wide-field FEL-MIP systems**

*Corresponding author Christoph.Krafft@leibniz-ipht.de

This comparison highlights the substantial increase in FOV and the potential for faster measurements with the wide-field FEL-MIP system, while also noting practical considerations such as the need for high laser power and the limited accessibility of FEL sources. In addition to Table 2, a previously published comparison of OPO- and QCL-based wide-field MIP systems provides a broader comparison of laser-specific parameters, including mid-IR output intensity, repetition rate, pulse duration, and achievable camera speeds[20]. This dataset highlights how differences in mid-IR laser characteristics influence imaging throughput and overall performance across wide-field MIP modalities.

The diameter of 45 μm in our wide-field QCL-MIP microscope has recently been expanded in related wide-field QCL-MIP approaches to a diameter of 63 μm with quantitative phase imaging[36], and to diameter of 100 μm with intensity diffraction tomography[37]. The authors of this study also noted[37] that the low mid-IR pulse energy of QCLs (few nJ) is the bottleneck of the FOV which requires future advances on the laser source side, such as pulsed mid-IR optical parametric oscillators (OPO). A high pulse energy OPO source (~10 μJ) with a repetition rate of 1 kHz was already coupled to a wide-field MIP microscope[20], however limited to the high wavenumber range 2800-3250 cm$^{-1}$. The FEL source which was applied here can emit up to 1 μJ per pulse at some wavelengths. As the repetition rate of the FEL was fixed at 13 MHz, this would give 13 W average power (here 0.15 μJ giving ca. 2 W average power in the fingerprint range, see table 1) which harbors the risk of damaging the samples. An off-axis parabolic mirror with longer focal length (see Fig. 5) would be required to further expand the IR beam. FOV can also be extended by implementation of a mosaic mode, that means collection of an array of FOVs and reconstruction of the full FOV by image stitching[19].

*Corresponding author Christoph.Krafft@leibniz-ipht.de

Additionally, the FEL offered tunability from 2000 down to 40 $cm^{-1}$ whereas the four chips of the QCL encompass the spectral range from 3000 to 2700 $cm^{-1}$ and 1800 to 934 $cm^{-1}$. However, FEL-MIP imaging below 1300 $cm^{-1}$ was impaired due to absorption of the $CaF_2$ substrate causing unwanted background. Other IR transparent substrates, such as ZnSe or float-zone silicon, might be better options in future work. Shorter IR pulse lengths of FEL in the ps range offers a better temporal resolution and more localized heating. The current FEL source studied here operates in the fingerprint region, but FELs in general offer a wide tuning range extending to the high-wavenumber region, along with variable pulse properties like pulse energy (<100 mJ) and lower repetition-rate modes, making them versatile for diverse photothermal imaging applications[38,39]. Disadvantages of FEL radiation are broader bandwidth near 16 $cm^{-1}$ compared to less than 1 $cm^{-1}$ bandwidth of IR pulses for better spectral resolution, and the power output of the FEL beam fluctuates more over time compared to the QCL output.

Both wide-field FEL- and QCL-MIP setups employed a counter-propagation mode. The photothermal images revealed a subtle enhancement at the PS bead edges, suggesting a possible contribution from heat diffusion. This observation could imply that the present microsecond probe pulse can possibly capture thermal spreading beyond the bead's surface. Future studies will utilize shorter (ns time scale), visible probe pulses with matching repetition rates to the FEL to reduce diffusion effects[25] and attain a more precise temporal and spatial resolution of the photothermal response. The commercial scanning OPTIR utilizes a Cassegrain objective in reflection mode, and the IR radiation excites a larger volume than the volume of the visible probe light. Here, the photothermal modulation directly changes the scattering field, and the modulation of the sample refractive index dominates. The OPTIR system exhibited more intense signals at the bead's center, whereas weaker signals were observed towards its edges, a phenomenon likely attributed to the bead's partial transparency, low scattering, and lens-like focusing properties. Our hyperspectral data (Fig. 2 and Fig. 6) agree with a recent study on PS beads with wide-field photothermal heterodyne imaging[40].

*Corresponding author Christoph.Krafft@leibniz-ipht.de

The effect of camera speed on SNR was previously demonstrated with polymethyl methacrylate samples[40]. SNR increased with frame averaging, as indicated by background suppressed mid-IR imaging[41]. The fitting curves of experimental SNR versus time for PS beads in supplementary Fig. S3 are broadly consistent indicating random electronic noise based on the central limit theorem. Supplementary Fig. S4 demonstrate how the SNR was calculated. In the FEL-MIP, the faster rise in SNR with frame averaging at high camera speed (200 fps) suggests that the camera speed can be increased further to achieve a higher SNR with lower frame averaging time. Unlike QCL, which emit IR pulses at approximately 1 kHz to 2 MHz, the FEL operates at 13 MHz; consequently, even at a high camera acquisition rate of 1000 fps, each frame would integrate roughly 13,000 IR pulses[42,43].

*Corresponding author Christoph.Krafft@leibniz-ipht.de

For the $CH_2$ deformation vibration near 1460 cm$^{-1}$ in tissue sections and cells, the QCL-MIP provided a modest contrast. The reason for this behavior is that the power is not evenly distributed by QCL lasers across the entire wavelength tuning range. In comparison to the amide I range near 1660 cm$^{-1}$, the power at 1460 cm$^{-1}$ range is significantly lower leading to weaker contrast which needs to be compensated by power scaling for reconstruction of IR spectra from hyperspectral images. Implementing a high-speed camera in FEL-MIP would allow wide-field image acquisition at higher time resolution and enable advances in many disciplines such as neuroscience to investigate the dynamics of chemical composition in functional brain tissues, or to map neurotransmitters, immune mediators and other relevant biomolecules involved in pathogenic processes[44,45]. Furthermore, the SNR of the FEL-MIP can also be enhanced by utilizing nanosecond pulsed visible probe lasers and CMOS sensors with a higher full well capacity[43]. Using a highly sensitive fluorescence detected photothermal infrared approach[46], the high average power of the FEL can be employed to further extend the field of view. Because it does not directly measure photon scattering or phase change[47] this fluorescence variant can reduce heat dissipation effects in photothermal contrast without using higher energy short pulsed mid-IR pump radiation.

26*Corresponding author Christoph.Krafft@leibniz-ipht.de

## Conclusion

A FEL was used for the first time as IR pump source for wide-field MIP imaging, and the FOV was enlarged by factor 20 compared to a QCL pump source. Due to the high repetition rate of 13 MHz the full power could not be utilized, and the power had to be reduced by a chopper and attenuators. A matching ns probe laser with MHz repetition rate and a camera with high frame rate and well capacity would contribute to further enlarged FOV. Furthermore, emerging approaches such as fluorescence-detected mid-IR imaging may provide complementary pathways to further improve sensitivity and expand the capabilities of FEL based mid-IR photothermal techniques for use in materials analysis and biomedicine.

## Materials and Methods

### Instrumentation

The wide-field QCL- and FEL-MIP microscope employed the same detection path. The function generator (9214+ pulse generator, Quantum Composers) operated at 50 kHz or 200 kHz (channel A) to externally trigger a blue LED (UHP-T-450- SR, Prizmatrix) with a central wavelength of 450 nm. A blue LED was chosen because its wavelength provides higher lateral resolution in the visible probe images than red wavelength, while its partially coherent emission minimizes speckle noise compared with coherent laser illumination. The LED was focused using a plano-covex lens with a 150 mm focal length (FL) and a plate beam splitter (35 mm×35 mm 50R/50T, VIS Plate Beamsplitter, Edmund optics) at the rear focal plane of the microscope objective (EC EPI Plan 40 /0.6 NA, Zeiss). A neutral density (ND) filter adjusts the LED intensity. The objective was fixed to a linear stage with a 12 mm travel (MT1/M-Z9, Thorlabs and KDC101, Thorlabs). A tube lens 1 (Zeiss tube lens, Edmund Optics) focused the reflected light from the sample onto the 24.5 Megapixels CMOS camera (MX245MG-SY-X4G3-FF, Ximea). With a 9, 4.5, or 2.25 ms exposure time for each frame, images were collected at 100, 200 or 400 fps. Considering the CMOS sensor size of 14.5×12.6 mm$^2$ and a 40× magnification/0.6 NA objective lens, the camera records visible images of 325 μm×280 μm areas. The CMOS pixel size of 2.74×2.74 μm$^2$ corresponds to an image area of ca. 70×70 nm$^2$ without binning and ca. 140×140 nm$^2$ after 2×2 binning. The images were stored in a buffer memory during the acquisition of frames. After the acquisition, they were transferred to a solid-state drive (SSD) via a high-speed Firefly interface. The difference of the mean hot and cold frames generated the photothermal contrast using Labview.

*Corresponding author Christoph.Krafft@leibniz-ipht.de

Wide-field QCL-MIP was coupled to a mid-IR QCL laser (Mircat 2400, Daylight Solutions). The QCL was triggered at 50 kHz or 200 kHz by a function generator, and the laser was switched on and off (20, 100 or 200 Hz at 50% duty cycle) using the duty cycle mode of the function generator. To focus and defocus the IR beam onto the sample, a ZnSe plano-convex lens with a 15 mm focal length (LA7477-E4, Thorlabs) was installed on the 12 mm linear stage. Initially, the response time of IR signals to an external trigger was assessed by inserting a CaF$_2$ beam splitter (BSW510, Thorlabs) in the beam path followed by an MCT detector (QM-10.6, Vigo Photonics). The $CaF_2$ beam was removed to avoid power loss during image acquisition.

*Corresponding author Christoph.Krafft@leibniz-ipht.de

To enable continuous acquisition of hot and cold frames in wide-field FEL-MIP, the IR beam (13 MHz) was modulated using a mechanical chopper (MC1F2, Thorlabs) positioned between two parabolic mirrors (MPD229-M01 and MPD239-M01, Thorlabs) to reduce the beam diameter. The chopper controller (MC2000B, Thorlabs) was set to a frequency of 50 Hz with a 50% duty cycle, chosen to match the camera frame rate and ensure synchronized capture of alternating FEL-on and FEL-off frames. The function generator acts as the master clock for the chopper, camera (90% duty cycle), and LED (50 kHz - 5% duty cycle). For precise focusing of the FEL beam onto the sample plane, we used a three-mirror configuration (Fig. 5) in which the beam from mirror M2 is directed onto an off-axis parabolic (OAP) mirror and then reflected by mirror M1 onto the sample plane. The OAP (MPD189-M01, Thorlabs) and mirror M1 (PF10-03-M02, Thorlabs) are positioned on a linear stage with a 12 mm travel range, facilitating precise adjustment of the gap between the OAP and the sample plane. This arrangement allows control of the beam focus: increasing the focal length of the OAP enlarges the illumination diameter at the sample, with simultaneous adjustment of M1 and the OAP ensuring that the FEL is precisely focused onto the sample plane for optimal wide-field excitation. Two step power attenuators were employed (infrared step attenuators, Lasnix); one was manually operated on the optical table and the other was mounted on the

29*Corresponding author Christoph.Krafft@leibniz-ipht.de

diagnostics table of the FELBE user laboratory at Helmholtz-Zentrum Dresden-Rossendorf (HZDR). The FEL power was measured using an Ophir 3A-P-THz thermal sensor, and the FEL spectrum was measured using an Acton Research SpectroPro-300 grating spectrometer.

**Sample preparation**

Commercial 10 µm polystyrene beads were suspended in distilled water. A few µl PS suspension was pipetted onto a 1 mm thick $CaF_2$ window (VUV grade). After drying, isolated and aggregated beads were evident. Lung tissue was taken from a C3HeB/FeJ mouse aerosol infected with *Mycobacterium tuberculosis* H37Rv for 30 days following bedaquiline treatment. After 44 days, the lung was extracted and fixed for 48 hours in a 4% paraformaldehyde solution in phosphate buffer. The lung was then incubated for 30 to 60 minutes in successive 5%, 10%, 15%, and 20% saccharose solutions. A Tissue-Tek cassette containing 20% saccharose was used to preserve the lung specimen. The tissue cassette was kept at -80°C after being snap-frozen in cold isopentane. The tissue segment was mounted onto a $CaF_2$ slide that was 1 mm thick after being cut into 12 $\mu$m. Cryosections of mouse brain and human larynx carcinoma were prepared at the University Hospital Jena at 10 µm thickness and transferred on 1.5 mm thick $CaF_2$ windows (VUV grade). THP-1 cells were cultured using standard protocols, fixed by formalin and stored in a PBS buffer. A few µl THP-1 cell suspension was transferred onto a 200 µm thick CaF2 window (VUV grade) and dried. Buffer crystals were re-dissolved by pipetting a few µl distilled water and removing the water droplet after a few seconds.

*Corresponding author Christoph.Krafft@leibniz-ipht.de

**Image processing**

Ximea CamTool and OPTIR studio were used to capture pictures for widefield and commercial microscopes, respectively. The following software was used to process the images (TIF) from the CMOS camera and OPTIR: LabVIEW 2024 for averaging images, and ImageJ and Python for further image and data processing.

**Acknowledgements**

This work is supported by Federal Ministry of Research, Technology and Space (BMFTR) within the funding programs QUANCER (13N16444) and Leibniz Center for Photonics in Infection Research (13N15464), and the EIC (European and Innovation Council) under the Horizon Europe program in the PATHFINDEROPEN-01 project TROPHY (grant no. 101047137).

**Author Contributions**

Conceptualization and methodology, C.K. and A.T.R.; investigation and formal analysis, A.T.R., S.A., A.S., J.M.K.; resources, J.M.K., N.R., O.G.L., J.P. and C.K.; software, A.T.R. and S.A.; writing—original draft, A.T.R. and C.K.; writing—review & editing, A.T.R., J.M.K. A.M., O.G.L. and C.K.; supervision, C.K. and J.P.; funding acquisition, O.G.L., C.K. and J.P. All authors read and agreed to the published version of the manuscript.

*Corresponding author Christoph.Krafft@leibniz-ipht.de

**Conflict of interest**

The authors declare no competing interests.

**Supporting information**

**Figure S1: Average IR power measured over the wavenumber range of 960 to 1800 cm$^{-1}$ using a power meter with a step size of 5 cm$^{-1}$.**

**Figure S2: Hyperspectral imaging of PS beads**

**Figure S3: Effect of duration of averaging photothermal images from polystyrene beads on SNR.**

**Figure S4: Photothermal image of PS beads at 1450 cm$^{-1}$, showing the locations where SNR was computed**

*Corresponding author Christoph.Krafft@leibniz-ipht.de

*Corresponding author Christoph.Krafft@leibniz-ipht.de

*Corresponding author Christoph.Krafft@leibniz-ipht.de

*Corresponding author Christoph.Krafft@leibniz-ipht.de